\documentclass[twocolumn,showpacs,preprintnumbers,amsmath,amssymb,pre]{revtex4}

\include{ams}
\usepackage{amsmath,amssymb,amsthm}
\usepackage{dcolumn}
\usepackage{bm}
\usepackage{graphicx}
\begin{document}
\title{Energy distributions of field emitted electrons from
carbon nanosheets: manifestation of the quantum size effect}

\author{V.L. Katkov}\email{katkov@theor.jinr.ru}

\author{V.A. Osipov}\email{osipov@theor.jinr.ru}

\affiliation{Bogoliubov Laboratory of Theoretical Physics, Joint Institute for Nuclear Research,
141980 Dubna, Moscow region, Russia}

\begin{abstract}
We emphasize the importance of experiments with voltage dependent
field emission energy distribution analysis
in carbon nanosheets. Our analysis shows the crucial influence
of the band structure on the energy distribution of field emitted
electrons in few-layer graphene. In addition to the main peak we found
characteristic sub-peaks in the energy distribution. Their
positions strongly depend on the number of layers and the
inter-layer interaction.
The discovery of these peaks in field
emission experiments from carbon nanosheets would be a clear
manifestation of the quantum size effect in these new materials.
\end{abstract}


\maketitle


Recently, freestanding carbon nanosheets (CNSs) have been synthesized
on a variety of substrates by radio frequency plasma enhanced chemical vapor
deposition~\cite{wang1,wang2}. The sheets are consisting of several graphene
layers and stand roughly vertical to the substrate. It has been found that CNSs have
good field emission characteristics with promising applications in vacuum microelectronic
devices~\cite{fecns1, fecns2, fecns3, Malesevic}. High emission total current at low threshold field
enables using CNSs as an effective cold cathode material.

Until now only the current-voltage characterization was used
in studies of CNSs. At the same time, voltage dependent
field emission energy distribution (V-FEED) analysis is known as a
powerful experimental method to interrogate the field emission. As
compared to classical I-V characterization, V-FEED analysis can
provide more information related to both inherent properties of
the emitter and to the basic tunneling process~\cite{Gadzuk}. In
particular, in single-walled carbon nanotubes (CNTs) the FEED has
shown characteristic peaks originated from the stationary waves in
the cylindrical part of the nanotube~\cite{tubes}. Their number
and sharpness were found to increase with the length of the tubes.
Notice that short periodic variations were also observed in the
thickness-dependent field emission current from ultrathin metal
films (UMF)~\cite{stark1}. The calculated electron energy
distribution curve characteristic of UMF was found to have "steps"
which correspond with the quantized "normal"
energies~\cite{stark2}. The resonant-tunneling peaks with specific
microscopic tunneling mechanisms were also observed in field emission
from nanostructured semiconductor cathodes~\cite{Johnson}.
A different example of the quantum size
effect in CNTs, which originates from the intrinsic properties of
the energy band structure, was revealed in field
emission~\cite{liang}. It is reasonable to expect manifestation of
quantum size effects in subnanometer CNSs.

In this Letter, we calculate the FEED of electrons from CNSs. For
this purpose, we take into account the energy band structure of
few-layer carbon systems resulting from the tight-binding
approach. Both the field emission current (FEC) and the FEED are calculated
by using the independent channel method suggested recently in
Ref.~\cite{Katkov}. Our analysis clearly shows that the FEC only
measurements give incomplete information. We found
that the FEED enables determination of the number of layers in few-layer
graphene as well as direct verification of the high sensitivity
of the band structure to the number of layers in few-layer graphene
reported recently in Ref.~\cite{Latil}.


Let us consider the graphene layer in the presence of
the external electric field $F$ directed along the $z$-axis
(see Fig.1).
\begin{figure}
\centering
\includegraphics[width=4cm,  angle=0]{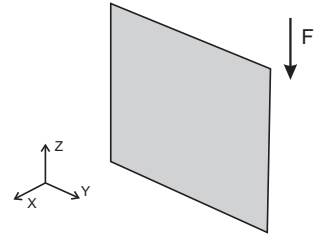}
\caption{The location of a graphene sheet with respect to the electric
field.} \label{fig.1}
\end{figure}
The emitted current density takes the following form:
\begin{equation}\label{main}
  j^{out} = \frac{2e}{h^{3}}\int d p_x
  \int d p_y \int
  f(\varepsilon)\upsilon_g D(\varepsilon, p_x, p_y)
  d p_z,
\end{equation}
where $e$ is the electric charge, $h=2\pi \hbar$ the Planck
constant, $\varepsilon$ the  energy, $\textbf{p}$ momentum,
$f(\varepsilon)=[\exp(\varepsilon/kT)+1]^{-1}$ the Fermi-Dirac
distribution function, $D(\varepsilon, p_x, p_y)$ the transmission
probability of an electron through a potential barrier, and
$\upsilon_g =\partial\varepsilon/\partial p_z$ the group velocity.
The integrals are over the first Brillouin zone with account
taken of the positivity of $\upsilon_g$.

For a two-dimensional (2D) structure, one can use the relation
$\int f(p_x) d p_x = f(0)h/l_x$. Moreover, when a graphene sheet
has the finite size in the $y$-direction, $p_y$ is quantized.
Therefore, the current density in Eq. (\ref{main}) can be written as
\begin{equation}\label{off}
j^{out} = \frac{2e}{h l_x
l_y}\sum_q\int_{\varepsilon_{min}^q}^{\varepsilon_{max}^q}
f(\varepsilon^q)D(\varepsilon^q)d \varepsilon^q.
\end{equation}
where the Fermi energy is chosen to be zero.
Limits $\varepsilon_ {max}^q $ and $\varepsilon_ {min}^q$ come from
the explicit form of the band structure.

We suggest that the transmission probability is given by the
WKB approximation in the form~\cite{Gadzuk}
\begin{equation}\label{DD}
D(\varepsilon) =
\exp\left[-\frac{\zeta(\phi-\varepsilon)^{3/2}\upsilon(y)}{F}\right],
\end{equation}
where $\zeta = 8 \pi(2 m)^{1/2}/3eh$, $y = (e
F/4\pi\varepsilon_0 )^{1/2}/\phi$,
$\phi$ is the work function, $\varepsilon_0$ the dielectric
constant, $m$ the electron mass.
The function $\upsilon(y)$
describes a deviation of the barrier from the triangle form due to
image effects and can be approximated as $\upsilon(y) \approx 1 -
y^{1.69}$ (see Ref.~\cite{Haw}).


The band structure of graphene multilayers has been obtained
within the tight-binding approach in Ref.~\cite{t-b}. Besides, an
approximation to the dispersion relation  can be found from
Slonzewski-Weiss-McClure (SWMcC) model for graphite with Bernal
stacking \cite{SW,McClure}. SWMcC model describes the wave-vector
dependence of electron energy in the vicinity of the $HKH$ edge of
the Brillouin zone (see Fig.2).
\begin{figure}
\centering
\includegraphics[width=3cm,  angle=0]{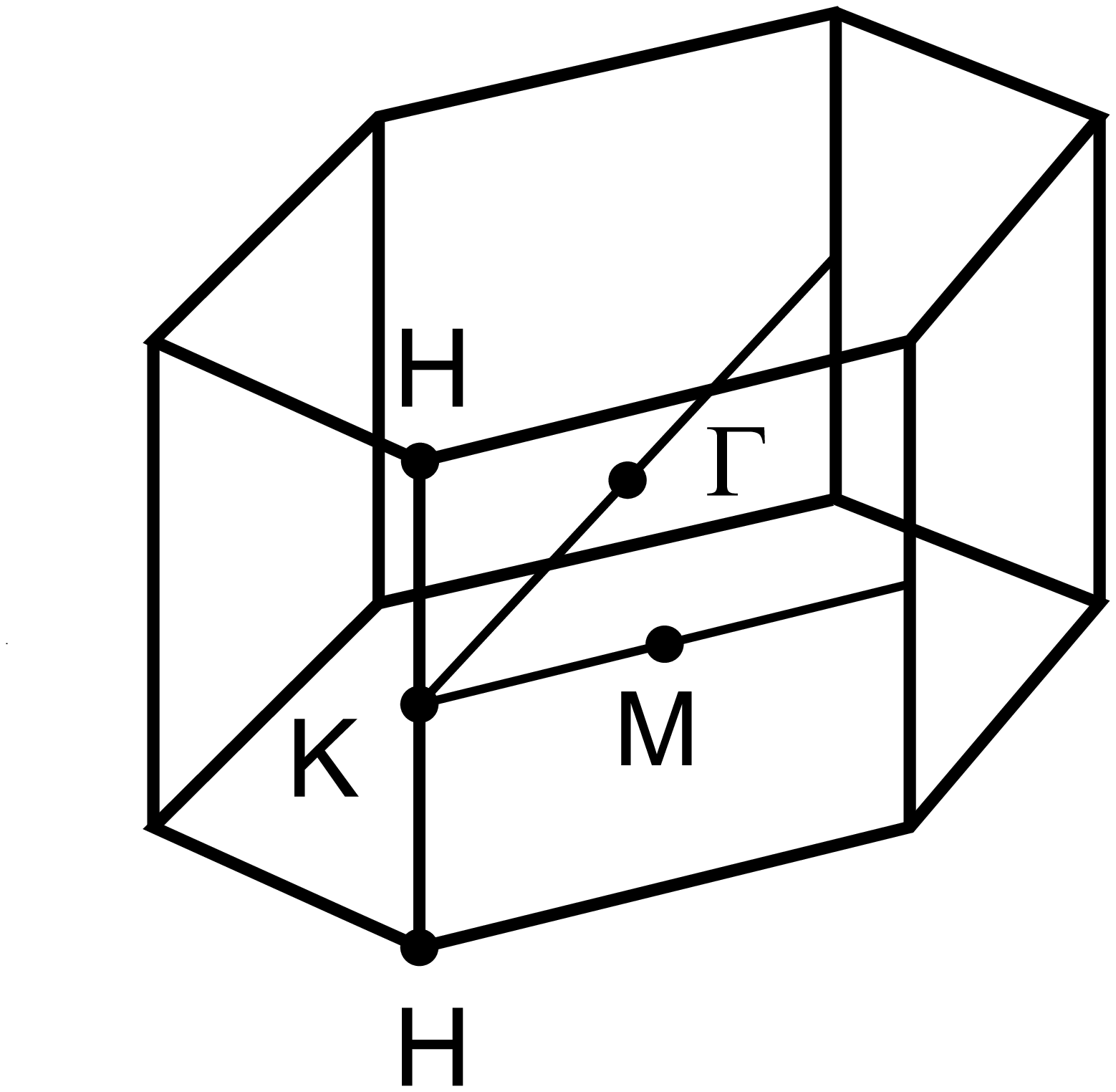}
\caption{The Brillouin zone of graphite.} \label{3}
\end{figure}
The electron energy spectrum is obtained from the equation
\begin{equation}
\det|H-\varepsilon|=0, \label{spectrum}
\end{equation}
where
\begin{equation}
H=\left(\begin{array}{cccc}E_1& 0 & H_{13} & H_{13}^{*}\\
0& E_{2} & H_{23} & -H_{23}^{*}\\
H_{13}^{*}& H_{23}^{*} & E_3 & H_{33}\\
H_{13}& -H_{23} & H_{33}^{*}& E_3\\
\end{array}\right)
\end{equation}
and
\begin{equation}
\begin{array}{c}
E_1 = \Delta + \gamma_1\Gamma +\frac{1}{2}\gamma_5\Gamma^2,\\
E_2 = \Delta - \gamma_1\Gamma + \frac{1}{2}\gamma_5\Gamma^2,\\
E_3 = \frac{1}{2}\gamma_2\Gamma^2,\\
H_{13} = \frac{1}{\sqrt{2}}(-\gamma_0 + \gamma_4\Gamma)\exp(i\alpha)\sigma,\\
H_{23} = \frac{1}{\sqrt{2}}(\gamma_0 + \gamma_4\Gamma)\exp(i\alpha)\sigma,\\
H_{33} = \gamma_3 \Gamma\exp(i \alpha)\sigma\\
\end{array}\label{p}
\end{equation}
with $\Gamma = 2 \cos(k_{\perp} c)$, $\sigma = k_{||}\sqrt{3}/2 a
 = p_{||}\upsilon_f/\gamma_1$, $k_{\perp}$ being
the wavevector projection onto the direction $HKH$, $k_{||}$
the modulus of the wavevector in the $yz$-plain, $\alpha$
the angle between $\textbf{k}_{||}$ and the
direction $\Gamma K$, $c$ the distance between nearest neighbour
layers, $a$ the lattice constant, $\textbf{p}_{||}$ the
momentum in the $yz$-plain, and $\upsilon_F$ the
Fermi velocity. Parameters $\gamma_i$ describe interactions between atoms and
$\Delta$ is the energy difference between two sublattices
in each graphene layer. In Ref.~\cite{DrH} graphite parameters were
estimated as $\gamma_0 = 3.16 ~eV, \gamma_1 = 0.39 ~eV, \gamma_2 =
-0.020 ~eV, \gamma_3 = 0.315 ~eV, \gamma_4 = -0.044 ~eV, \gamma_5
= 0.038 ~eV$ and $\Delta = -0.008 ~eV.$
The spectrum of few-layer graphene can be obtained from Eq.~(\ref{spectrum})
by replacing $\Gamma$ by
\begin{equation}
\Gamma^n = 2\cos\left(\frac{\pi n
}{N+1}\right),~~ n = 1\dots N, \label{gamma}
\end{equation}
where $N$ is the number of layers. For graphene bilayer $N=2$ and
$\gamma_2=\gamma_5=0$ so that Eq. (\ref{spectrum}) with account
taken of  Eq. (\ref{gamma}) gives the result of
Ref.~\cite{Mikitik} while for $N=1$  (only $\gamma_0$ differs from
zero) it reproduces the known tight-binding spectrum of graphene.

As a first approximation one can neglect all interactions except between the
nearest-neighbor atoms in the same layer and between
A-type atoms between adjacent layers (which are on top of
each other), i.e. all parameters except for $\gamma_0$ and $\gamma_1$
are putted to be zero. Then the spectra of multilayers can be approximated
by
\begin{equation}\label{Seq1}
\varepsilon^n_{c,v} = \pm\left(\sqrt{(\gamma_1^n/2)^2+p^2_{||}\upsilon_F^2}-\gamma_1^n/2\right),
\end{equation}
where $\gamma_1^n = \gamma_1 \Gamma^n$.


The FEC and the FEED ($P(\varepsilon)$) are connected by (see,
e.g., Ref.~\cite{Gadzuk})
\begin{equation}\label{eq.2}
j^{out} =\int\limits_{-\infty}^{\infty} d\varepsilon P(\varepsilon).
\end{equation}
The explicit form of $P(\varepsilon)$ for few-layer graphene can be
found from Eq. (\ref {off}). Indeed, for layers of a large (infinite)
size the sum in Eq. (\ref{off}) can be replaced by the integral and,
correspondingly, one has to use $\varepsilon_{min}(p_y)$ and
$\varepsilon_{max}(p_y)$ instead of $\varepsilon_{min}^q$ and
$\varepsilon_{max}^q$. In our case, these $p_y$-dependent functions
can be easily calculated from Eq. (\ref{Seq1}). Finally, we have to
change the order of integration in Eq. (\ref{off}). The result is
\begin{equation}
 P(\varepsilon) =
\frac{2g}{\upsilon_F}
f(\varepsilon)D(\varepsilon)\sum\limits_{n=1}^{n=N}
\theta(\epsilon_n)\sqrt{|\varepsilon|\epsilon_n},\label{FEED}
\end{equation}
where $\epsilon_n=|\varepsilon|+\gamma^n_1$, $g =4 e/(h^2 N c)$,
and $\theta(\epsilon)$ is the Heaviside step function.

\begin{figure}
\centering
\includegraphics[width=8cm,  angle=0]{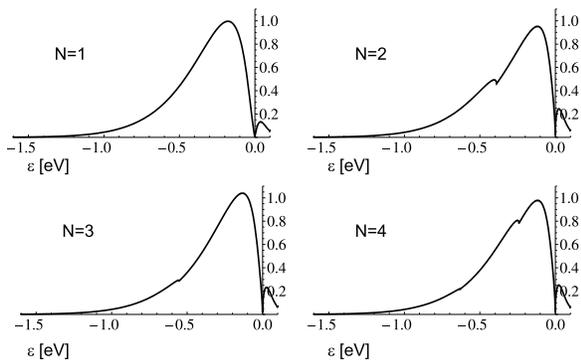}
\caption{Reduced FEED for one- to four-layer graphene.
All SWMcC parameters except for $\gamma_0$ and $\gamma_1$
are putted to be zero, $F = 4$ V/nm. The peak height for $N=1$
is chosen to be unity.} \label{fig.3}
\end{figure}

For graphene monolayer one gets
\begin{equation}
P^{mono}(\varepsilon) = \frac{2g}{\upsilon_F}f(\varepsilon)D(\varepsilon)|\varepsilon|.
\end{equation}
Within the Fowler-Nordheim (FN) approximation (weak fields and low temperatures)
the FEC is found to be
\begin{equation}\label{mono}
j^{mono} =  \frac{2gb}{\upsilon_F d^2},
\end{equation}
where $b =\exp\left(-\zeta\phi^{3/2} \upsilon(y)/F\right)$ and $d =
3\zeta\phi^{1/2} t(y)/2F$ with $t(y)\approx 1+0.1107y^{1.33}$ (see Ref.~\cite{Haw}).
For bilayer, the FEED is obtained as
\begin{equation}\label{Nbi}
P^{bi}(\varepsilon) = \frac{2g}{\upsilon_F}
f(\varepsilon)D(\varepsilon)
\left(\sqrt{|\varepsilon|\epsilon_1}
+ \theta(\epsilon_2)\sqrt{|\varepsilon|\epsilon_2}\right),
\end{equation}
and, correspondingly,
\begin{equation}\label{bi21}
j^{bi} = \frac{gb\gamma_1}{\upsilon_F d}\exp\left({\frac{d
\gamma_1}{2}}\right) K_1\left(\frac{d \gamma_1}{2}\right),
\end{equation}
where $K_1(x)$ is the MacDonald function. Notice that only the
first term in Eq. (\ref{Nbi}) is significant at weak fields. When
the interlayer interaction is weak ($\gamma_1 d\ll 1$)
Eq.(\ref{bi21}) passes into Eq.(\ref{mono}). For large $\gamma_1
d$ one gets
\begin{equation}
j^{bi} =\frac{gb}{\upsilon_F d^2}\sqrt{\pi \gamma_1 d}.
\end{equation}
Thus, we obtain a standard FN exponent while the preexponential factor
becomes proportional to $F^{3/2}$ instead of $F^2$ for the FN theory.
Fig.~\ref{fig.3} shows the FEED for different numbers of graphene layers.

\begin{figure}
\centering
\includegraphics[width=8cm,  angle=0]{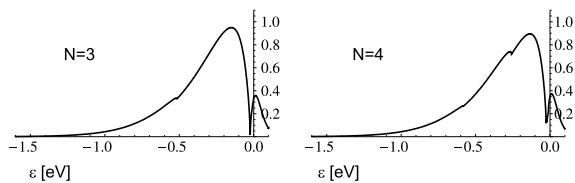}
\caption{Reduced FEED for three- and four-layer graphene. The whole set of SWMcC
parameters is used,  $F = 4$ V/nm.} \label{fig.4}
\end{figure}

Let us now take into account all possible interactions. For this purpose,
we use the whole set of SWMcC parameters and put $\alpha =0$.
The numerical results are presented in Fig.~\ref{fig.4}.
It should be stressed that for $N=1$ and $N=2$ the calculated
FEEDs are not sensitive to other interaction constants and curves
are found to be identical to those shown in Fig.~\ref{fig.3}. When
$N$ increases little shifts of the minima relative to the Fermi
energy are obtained. As is clearly seen, for $N>1$ FEEDs have
characteristic sub-peaks. The number of peaks and their positions
strongly depend on the number of layers and the interaction
constants, first of all, $\gamma_1$. There is a pronounced
depression in FEED at the Fermi energy which would be typical for
3D gapless semiconductors.

\begin{figure}
\centering
\includegraphics[width=6cm,  angle=0]{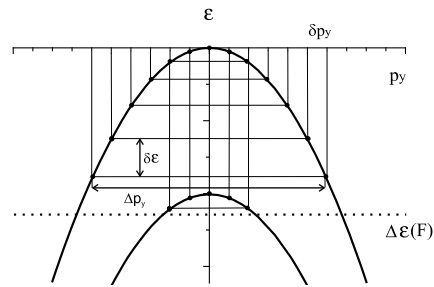}
\caption{Schematic illustration of the method of independent channels for
few-layer structures. Black points indicate peaks of emission channels.} \label{fig.5}
\end{figure}

Fig.~\ref{fig.5} gives a clear illustration of our results.
At room temperatures, the Fermi-Dirac distribution function
restricts the FEED above the Fermi energy, so that the
electron emission from the valence band dominates.
The transmission probability decays exponentially with
decreasing $\varepsilon$. Therefore, at fixed $F$ one can estimate
the energy range of emitted electrons as
\begin{equation}
\triangle \varepsilon\approx\frac{2}{3\zeta\phi^{1/2}}F,
\end{equation}
In accordance with Fig.~\ref{fig.5} the number of emitting
channels $m$ at the energy $\varepsilon$ is defined as $m=
[\triangle p_y(\varepsilon)/\delta p_y]$ where the brackets $[~]$
indicate integer part. Generally, $P(\varepsilon) \sim
C(\varepsilon)f(\varepsilon)D(\varepsilon)$ where
$C(\varepsilon)=m/l_y$ is the density of emitting channels. In
CNSs $l_y$ is large enough and $C(\varepsilon)\rightarrow\triangle
p_y(\varepsilon)/h$. In our case, $\triangle
p_y(\varepsilon)\rightarrow 0$ at $|\varepsilon|\rightarrow 0$ so
that $P(\varepsilon)\rightarrow 0$ (see Fig.~\ref{fig.3}).
Interestingly that taken into account Eq.~(\ref{Seq1}) one can
easily calculate $C(\varepsilon)$ and, correspondingly,
$P(\varepsilon)$ for any $N$ without integrations. The shape of
the FEED in Figs.~\ref{fig.3} and \ref{fig.4} directly depends on
the density of emission channels. When $\varepsilon$ riches the
top of the next branch of the spectrum this branch becomes
"switched-on" thus resulting in a distinctive point in the FEED.
Evidently, the closer a position of the branch to $\triangle
\varepsilon$ the less pronounced is an additional peak in
$P(\varepsilon)$.

An important difference from the emission of single-walled carbon
nanotubes should be mentioned. The diameters of  CNTs are very
small thus resulting in a set of discrete channels. For metallic
CNTs there also exists at least one emitting channel at
$|\varepsilon|\rightarrow 0$. However, as distinct from CNSs the
density $C(\varepsilon)$ tends to a constant value at
$\varepsilon\rightarrow 0$ and, correspondingly, the FEED exhibits
behavior typical for conventional metallic emitters without any
minimum near the Fermi energy. On the contrary, in semiconducting
CNTs the FEED has a characteristic gap at the Fermi energy (cf.
Ref.~\cite{liang2}).

\begin{figure}
\centering
\includegraphics[width=8cm,  angle=0]{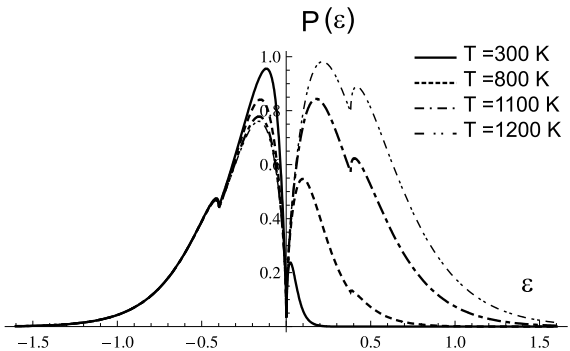}
\caption{FEED for bilayer at different temperatures. The whole set of SWMcC
parameters is used,  $F = 4$ V/nm.} \label{fig.20}
\end{figure}

\begin{figure}
\centering
\includegraphics[width=5cm,  angle=0]{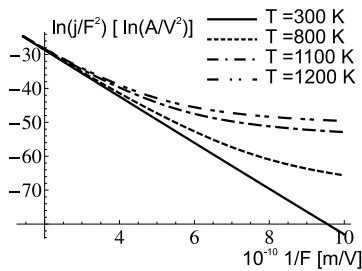}
\caption{FEC for bilayer in Fowler-Nordheim coordinates at different temperatures.} \label{fig.19}
\end{figure}

Similar arguments are valid for the emission from the conduction
band where, however, the limiting role of the Fermi-Dirac
distribution is of decisive importance.
When the temperature grows, the electrons from the conduction band
become involved in the emission. As is seen in Fig.~\ref{fig.20},
the regime of the so-called thermal field emission occurs at high
temperatures of emitters.
Notice that there is a rather symmetrical
behavior of curves in both bands, which is valid for all considered
few-layer structures. Fig.~\ref{fig.19} shows an influence of the
thermal field emission on the behavior of emission current in the FN coordinates.
The curves are found to be practically identical for any $N$.
At high temperatures one can see marked deviations from the standard FN plot
in the region of weak fields. Notice that similar deviations were observed
in experiments with as-received CNSs (see Fig. 9 in Ref.~\cite{fecns1}).


In conclusion, we have calculated the FEED for few-layer graphene
films and found the presence of characteristic sub-peaks
originated from involving in the emission process additional
branches in the energy spectrum of layered structures. Since
the peak positions are directly determined by the number of
layers the discovery of such peaks in the FEED would be a clear
manifestation of the quantum size effect. Therefore,
the experimental studies of the FEED for CNSs are very relevant.
Furthermore,
the FEED analysis gives a new experimental tool to estimate the
inter-layer interaction constants (along with Raman scattering in
Ref.~\cite{Malard} and photoemission methods in Ref.~\cite{Ohta}) and provides
important information on the concrete types of emitting CNSs as well as
allows one to identify the number of layers in emitting CNSs.
For example, the absence of sub-peaks would indicate that the emission
occurs from monolayer graphene. In addition, the emitter temperature is taken
into account in the FEED via the Fermi-Dirac distribution function and can be
determined by the half-width and the relative height of shapes in
the conduction band. Finally, using an approach suggested in
Ref.~\cite{tfem1} one can measure the resistivity of CNSs.

This work has been supported by the Russian Foundation for Basic
Research under grant No. 08-02-01027.



\begin{thebibliography}{99}

\bibitem{wang1} J. J. Wang et al., Appl. Phys. Lett. \textbf{85}, 1265
(2004).
\bibitem{wang2} J. J. Wang et al., Carbon \textbf{42}, 2867
(2004).
\bibitem{fecns1} M. Bagge-Hansen et al., J. Appl. Phys. \textbf{103}, 014311
(2008).
\bibitem{fecns2} Kun Hou et al., Appl. Phys. Lett. \textbf{92}, 133112
(2008).
\bibitem{fecns3} Goki Eda et al., Appl. Phys. Lett. \textbf{93}, 233502
(2008).
\bibitem{Malesevic} A. Malesevic et al., J. Appl. Phys. \textbf{104}, 084301 (2008).
\bibitem{Gadzuk} J. W. Gadzuk and E.W. Plummer, Rev. Mod. Phys. \textbf{45}, 487
(1973).
\bibitem{tubes}  A. Mayer, N. M. Miskovsky, and P. H. Cutler, J. Phys.: Condens. Matter \textbf{15}, R177
(2003).
\bibitem{stark1} D. Stark and P. Zwicknagl, J. Appl. Phys. \textbf{21}, 397406
(1980).
\bibitem{stark2} J. K. Wysockia and D. Stark, Surf. Sci. \textbf{247}, 402
(1991).
\bibitem{Johnson} S. Johnson, U. Z\"{u}licke, and A. Markwitz, J. Appl. Phys. \textbf{101}, 123712
(2007).
\bibitem{liang} S.-D. Liang et al., Appl. Phys. Lett. \textbf{85}, 813 (2004).
\bibitem{Katkov} V. L. Katkov and V. A. Osipov, J.Phys.: Condens. Matter \textbf{20}, 035204
(2008).
\bibitem{Latil} S. Latil and L. Henrard, Phys. Rev. Lett. \textbf{97}, 036803 (2006).
\bibitem{Haw} P. W. Hawkes and E. Kasper, {\it Principles of Electron Optic}, (Academic Press, London, 1989), Vol. 2.
\bibitem{t-b} B. Partoens and F. M.Peeters, Phys. Rev. B \textbf{74}, 075404 (2006).
\bibitem{SW} J. C. Slonzewski and P. R. Weiss, Phys. Rev. B \textbf{109}, 272 (1958).
\bibitem{McClure} J. W. McClure, Phys. Rev. B \textbf{108}, 612 (1957).
\bibitem{DrH} M. S. Dresselhaus and G. Dresselhaus, Adv. Phys. \textbf{30}, 139 (1981).
\bibitem{Mikitik} G. P. Mikitik and Yu. V. Sharlai, Phys. Rev. B \textbf{77}, 113407 (2008).
\bibitem{liang2} S.-D. Liang et al., Phys. Rev. B \textbf{73}, 245301 (2006).
\bibitem{Malard} L. M. Malard et al., Phys. Rev. B \textbf{76}, 201401(R) (2007).
\bibitem{Ohta} T. Ohta et al., Phys. Rev. Lett. \textbf{98}, 206802 (2007).
\bibitem{tfem1} S. T. Purcell et al., Phys. Rev. Lett. \textbf{88}, 105502 (2002).


\end{thebibliography}
\end{document}